\documentclass[12pt]{article}
\usepackage{amsmath}
\usepackage{latexsym}
\usepackage{bbm}
\usepackage{amssymb}

\def\a{\alpha}

\def\b{\beta}

\def\d{\delta}

\def\k{\kappa}
\def\l{\lambda}

\def\o{\omega}

\def\D{\Delta}

\def\pa{\partial}

\def\and{{\rm and}}

\def\ie{{\it i.e.,} }

\def\IZ{{\mathbbm Z}}

\def\mbf{\mathbf}

\def\and{{\quad {\rm and} \quad}}

\def\A5{{AdS_5 \times S^5}}

\newcommand{\be}{\begin{equation}}
\newcommand{\bea}{\begin{eqnarray}}

\newcommand{\ee}{\end{equation}}
\newcommand{\eea}{\end{eqnarray}}

\begin{document}
\vspace*{-1.0in}
\thispagestyle{empty}
\begin{flushright}
CALT-TH-2022-030
\end{flushright}

\normalsize
\baselineskip = 18pt
\parskip = 6pt

\vspace{1.0in}

{\Large \begin{center}
{\bf Diffeomorphism Symmetry in Two Dimensions and Celestial Holography}
\end{center}}

\vspace{.25in}

\begin{center}
John H. Schwarz\footnote{jhs@theory.caltech.edu}
\\
\emph{Walter Burke Institute for Theoretical Physics\\
California Institute of Technology 452-48\\ Pasadena, CA  91125, USA}
\end{center}

\vspace{.25in}

\begin{center}
\textbf{Abstract}
\end{center}
\begin{quotation}
Two-dimensional diffeomorphism symmetry can be described by an operator algebra
extension of the well-known Virasoro algebra description of conformal symmetry.
Utilizing this extension, this note explains why the conformal symmetry that
appears in celestial holography
should not be extended to diffeomorphism symmetry, a possibility that
several authors have proposed. The description of the two-dimensional diffeomorphism
algebra presented here might be useful for other purposes.
\end{quotation}

\newpage


\pagenumbering{arabic}



\section{Introduction}

Unlike in higher dimensions, the conformal group in two dimensions is
infinite dimensional. In the case of Euclidean signature, which is
the main concern in this work, the generators consist of the product
of a holomorphic Virasoro algebra and its complex conjugate
antiholomorphic algebra. An enormous amount of work has
taken place over the past 50 years studying 2d conformal field theories.
It is a rich subject with many important applications.

Two-dimensional Euclidean space can be parametrized by a complex coordinate $z$
and its complex conjugate $\bar z$. By means of stereographic projection, these
coordinates can also be used to parametrize a sphere.
By definition, a conformal operator $\Phi (z, \bar z)$ has dimensions $(h, \bar h)$
provided that
$\Phi(z, \bar z) (dz)^h  (d\bar z)^{\bar h}$ is invariant under conformal
transformations. It is customary to define {\em dimension} $\D = h +\bar h$
and {\em spin} $s = h - \bar h$. The mode expansion of such an operator is
\be \label{rule}
\Phi (z, \bar z) = \sum_{m,n} \frac{\Phi_{m,n}}
{z^{m+h} \, {\bar z}^{n+\bar h}}.
\ee
The range of $m$ and $n$ are such that $m+h$ and $n+\bar h$ take all integer values.


Conformal transformations are generated by the operators
\be
T(z) = \sum_{m \in \IZ}\frac{ L_m}{z^{m+2}} \and
\overline T(\bar z) = \sum_{m \in \IZ} \frac{{\overline L}_m}{ {\bar z}^{m+2}}.
\ee
The modes satisfy the Virasoro algebra
\be
[L_m, L_n] = (m - n)L_{m+n}, \quad [{\overline L}_m, {\overline L}_n]
= (m - n){\overline L}_{m+n}, \quad
[L_m, {\overline L}_n] = 0.
\ee
No central extension is required.\footnote{See \cite{Barnich:2017ubf} for
the possible role of a central extension in the context of celestial holography.}
Conformal transformations of an operator $\Phi (z, \bar z)$
having dimensions $(h, \bar h)$ are generated by the Virasoro operators
\be
[ L_k , \Phi_{m,n}]
= ((h-1)k - m)\Phi_{m+k,n},
\ee
\be
[ {\overline L}_l , \Phi_{m,n}]
= ((\bar h -1)l -n ) \Phi_{m,n+l}.
\ee
These commutation relations
correspond to the operator-product formulas (as $z \to w$)
\be
T(z) \Phi (w, \bar w)\sim h \frac{\Phi (w, \bar w)}{(z-w)^2}
+ \frac{\pa \Phi (w, \bar w)}{z-w} ,
\ee
\be
\overline T(\bar z) \Phi (w, \bar w)
\sim \bar h \frac{\Phi (w, \bar w)}{(\bar z - \bar w)^2}
+ \frac{\bar\pa \Phi (w, \bar w)}{\bar z- \bar w} ,
\ee
where $\pa = \pa/\pa w$ and $\bar\pa = \pa/\pa \bar w$.

Conformal algebras can be described either by commutation
relations for modes of operators and or by the identification of the
singular terms in operator product expansions (OPEs), as we have just
demonstrated. The latter is
more succinct and elegant, and therefore it is usually preferred.
One advantage of considering the commutation relations of modes is
that this makes it easier to check consistency with Jacobi identities of the form
$[[A,B],C] + [[B,C],A] + [[C,A],B] =0$. There are cases where checking this
consistency is important, as will become clear later.
These Jacobi identities correspond to associativity of the operator algebra,
which can be investigated by introducing appropriate contour integrals.



One of the important precursors of celestial holography was the
famous BMS analysis in the early 1960s \cite{Bondi:1962px}\cite{Sachs:1962wk}\cite{Sachs:1962zza}.
These authors analyzed the symmetries of general relativity in
asymptotically flat spacetime (AFS) geometries. They
concluded that the usual translation generators, \ie momenta, are supplemented
by an infinite set of additional ones, which they called {\em supertranslations}.
Similarly, the Lorentz generators acquire an infinite
extension, which are called {\em superrotations}.
These conclusions were reached by analyzing the asymptotic behavior
of the spacetime metric in appropriately chosen coordinates, subject
to assumptions about the asymptotic falloff of various functions that appear in
the analysis.

Eventually it was realized that the
BMS algebra should be extended to include the entire infinite-dimensional
Virasoro algebra \cite{deBoer:2003vf}\cite{Barnich:2009se}\cite{Barnich:2010eb}.
This implies that the Lorentz symmetry together
with the superrotations can be interpreted as the infinite-dimensional conformal
symmetry of the two-dimensional celestial sphere.\footnote{Other
possibilities considered in the literature include various Carrollian groups.
See \cite{Duval:2014lpa} -- \cite{Bagchi:2022emh}, for example.}
Furthermore, there is compelling evidence for the existence of
a dual description of gravitational physics in asymptotically flat
four-dimensional spacetime in terms of a 2d conformal field theory on
the celestial sphere. Much of the pioneering work has been done by Strominger,
who has written a book on the subject \cite{Strominger:2017zoo}.
There are also a number of more recent reviews including
\cite{Raclariu:2021zjz}--\cite{McLoughlin:2022ljp}.

A basic premise of celestial holography is that bulk fields in four-dimensional
asymptotically flat spacetime map to operators on the celestial sphere.
Furthermore, spacetime scattering amplitudes map to correlation functions
of these operators on
the celestial sphere. Operator product expansions
of these operators on the celestial sphere
play a central role in the current research. They correspond to collinear
limits of the bulk fields.

In recent years, various authors have revisited the BMS analysis. By making slight changes
to the assumptions some of them have been led to conclude that the asymptotic
symmetry is larger than the conformal group. Some of these studies are
\cite{Campiglia:2014yka}--\cite{Chandrasekaran:2021vyu}.
In most cases they appear to conclude that the conformal symmetry of celestial holography,
Conf($S^2$), should be extended to diffeomorphism symmetry, Diff($S^2$).
It is difficult to decide which conclusion is correct
within the general relativity framework that is used in these analyses. Some
works, such as \cite{Flanagan:2019vbl}\cite{Krishnan:2022eff}, are careful to
distinguish between asymptotic spacetime symmetries and symmetries of the
celestial CFT, even though they must be closely related. For previous studies of
Diff($S^2$) and related algebras see \cite{Enriquez-Rojo:2021rtv}.

The goal of this paper is to answer this question by using
considerations that do not refer explicitly to a spacetime metric.
Specifically, the plan is to explore what is involved in extending
the conformal symmetry algebra of a 2d Euclidean
conformal field theory to diffeomorphism symmetry and what its physical
implications are in the context of celestial holography. We will
conclude that the conformal symmetry in celestial holography
should not be extended to diffeomorphism symmetry.

\section{The diffeomorphism algebra}

The key fact is that the conformal symmetry generator $T(z)$ describes a subset of
the modes of a diffeomorphism generator $J(z, \bar z)$. The complex
conjugate statement is that $\overline T (\bar z)$ describes a subset of the modes
of $\overline J (z, \bar z)$. The mode expansions of these operators have the form
given previously for operators of dimension $(2,1)$ and $(1,2)$, respectively. Thus,
\be
J(z, \bar z) = \sum_{mn} \frac{J_{m,n}}{z^{m+2} \, {\bar z}^{n+1}} \and
\overline J(z, \bar z) = \sum_{mn} \frac{{\overline J}_{m,n}}{z^{m+1}\, {\bar z}^{n+2}}
\quad m,n\in\IZ.
\ee
However, as we will see, this statement does not adequately describe their algebra.
The embedding of $T$ and $\overline T$ is given by
\be J_{m,0} = L_m \and {\overline J}_{0,n} = {\overline L}_n.
\ee

Aside from a possible central extension, the natural extension of the conformal
symmetry algebra to the diffeomorphism algebra is
\be
[J_{k,l}, J_{m,n}] = (k-m) J_{k+m, l+n}
\ee
\be
[{\overline J}_{k,l}, {\overline J}_{m,n}] = (l-n) {\overline J}_{k+m, l+n}
\ee
\be \label{JbarJ}
[J_{k,l}, {\overline J}_{m,n}] = l J_{k+m, l+n} - m {\overline J}_{k+m, l+n}.
\ee
The coefficients have been determined by setting various indices to zero and
implementing the correct conformal subalgebra. As we will show, these terms
are correct, but another term involving a new operator needs to be added.

The fact that this is essentially the correct algebra for the generators of
diffeomorphisms, also known as general coordinate transformations,
is shown by representing them by differential operators
\be
J_{k,l} \sim  z^{1-k}{\bar z}^{-l} \frac{\pa}{\pa z} \quad {\rm and} \quad
{\overline J}_{m,n} \sim z^{-m} {\bar z}^{1-n} \frac{\pa}{\pa \bar z} .
\ee
Operators of dimension $(h, \bar h)$ satisfy the same algebra as conjugate operators
of dimensions $(1-h, 1-\bar h)$. (The conjugate operator is called the ``shadow transform.")
That is why the $(2,1)$ and $(1,2)$ operators $J$ and $\overline J$
satisfy the same algebra as the $(-1, 0)$ and $(0, -1)$ differential operators
that transform the coordinates $z$ and $\bar z$. It is also why we used the $\sim$ symbol
here. These differential operators will not be used in the subsequent analysis.

If $\Phi$ is an operator with dimensions $(h, \bar h)$,
but not $J$ or $\overline J$, diffeomorphisms are encoded in the commutation relations
of its modes
\be \label{JPhi}
[J_{k,l}, \Phi_{m,n}] = ((h-1)k-m) \Phi_{k+m, l+n}
\ee
\be \label{barJPhi}
[{\overline J}_{k,l}, \Phi_{m,n}] = ((\bar h -1)l -n) \Phi_{k+m, l+n}.
\ee
These formulas play a crucial role in the subsequent analysis, so we should
say more about them.
The first formula is certainly correct for $l=0$, since $J_{k,0} =L_k$,
so any modification should be proportional to $l$. Similarly, any modification
of the second formula should be proportional to $k$. So let us be safe and
(temporarily) write
\be 
[J_{k,l}, \Phi_{m,n}] = ((h-1)k-m +\a l) \Phi_{k+m, l+n}
\ee
\be 
[{\overline J}_{k,l}, \Phi_{m,n}] = ((\bar h -1)l -n +\b k) \Phi_{k+m, l+n},
\ee
where $\a$ and $\b$ are allowed to depend on $h$ and $\bar h$.

Diffeomorphism invariance is not sufficient by itself to describe theories
containing operators with nonzero conformal spin. To demonstrate the issue
let us examine the Jacobi identity involving $J$, $\overline J$, and an operator
$\Phi$ that has dimensions $(h, \bar h)$. (There is no problem with the $JJ\Phi$ and
$\overline J \overline J \Phi$ Jacobi identities.) All of the requisite commutators
have been presented. Using them we find that
$$
[[J_{k,l}, {\overline J}_{m,n}], \Phi_{p,q} ]
+ [[{\overline J}_{m,n}, \Phi_{p,q} ], J_{k,l} ]
+ [[\Phi_{p,q}, J_{k,l}], {\overline J}_{m,n}]
$$
\be
=  [(h - \bar h) lm +\a ln -\b km]\, \Phi_{k+m+p, l+n+q},
\ee
which shows that the algebra is inconsistent as it stands. The only way to rectify it
is to set $\a =\b =0$ and to add an additional term to the $[J, \overline J]$ equation.

The appearance of the spin of $\Phi$,
$h - \bar h$, suggests the need to incorporate a spin operator in the formulas. For
this purpose let us introduce a dimension $(1,1)$ spin operator $S(z, \bar z)$, with the
commutation relation
\be \label{Srule}
[S_{k,l}, \Phi_{m,n}]= (h-\bar h) \Phi_{k+m, l+n}.
\ee
Then we add another term to eq. (\ref{JbarJ}) giving
\be
[J_{k,l}, {\overline J}_{m,n}] = l J_{k+m, l+n} - m {\overline J}_{k+m, l+n} - lm S_{k+m, l+n},
\ee
which repairs the Jacobi identity. Additional Jacobi identities require that
\be
[J_{k,l}, S_{m,n}] = -m S_{k+m,l+n},
\ee
\be
[{\overline J}_{k,l}, S_{m,n}] = -n S_{k+m,l+n},
\ee
\be
[S_{k,l}, S_{m,n}] = 0.
\ee

Note that $J$ and $\overline J$ are special and do not follow the
general rule for operators of dimension $(2,1)$ and $(1,2)$. In particular, eq.~(\ref{Srule})
does not apply to them. Instead, they transform $S$ in accordance with
their general rule in eqs.~(\ref{JPhi}) and (\ref{barJPhi}).
This conclusion was reached by studying Jacobi identities.

We can now convert the various commutation relations
of modes into equivalent operator product expansions.
Typically what happens when one evaluates a commutator of
two operators, such as $[A(z,\bar z), B(w, \bar w)]$ by inserting mode expansions for
each of them is that one encounters ill-defined series of the form
$\sum_m w^{m}/z^{m+1}$ times some operator $C(w)$. The ill-defined series,
as well as its derivatives and complex conjugate,
need to be interpreted. These series determine the poles in the corresponding
OPE, which carry all of the relevant information.
A convenient mnemonic that allows one to
convert between commutators of modes and singular OPEs
in either direction uses the correspondence
\be
\sum_m \frac{w^{m}}{z^{m+1}} \leftrightarrow \frac{1}{z-w},
\ee
as well as derivatives and complex conjugates of this correspondence.
This rule reproduces the results obtained by the contour integral
methods mentioned earlier.

We are now ready to re-express the complete algebra of the $J$,
$\overline J$, and $S$ symmetry operators, combining diffeomorphism
symmetry with the spin operator, as well as their action on
an arbitrary conformal operator $\Phi$ of dimensions $(h, \bar h)$.
The OPEs of various $\Phi$ operators among themselves
addresses the dynamics of specific theories and (with one exception
that will appear later) is beyond the scope of this note.

Using the procedure sketched above to deduce the OPEs from
commutation relations, we find the following\footnote{In
order to save space, we do not display the $w$ and $\bar w$ dependence
of the operators appearing on the right-hand side of the OPEs.}
\be
S(z, \bar z) S(w,\bar w) \sim 0.
\ee
\be
J(z, \bar z) S(w,\bar w) \sim \frac{1}{|z - w|^2}
\left( \frac{ S}{z-w} + \pa S \right)
\ee
\be
\overline J(z, \bar z) S(w,\bar w) \sim \frac{1}{|z - w|^2}
\left(\frac{S}{\bar z -\bar w}
+ \bar\pa S \right)
\ee
\be
J(z, \bar z) J(w,\bar w) \sim \frac{1}{|z - w|^2}
\left(\frac{2 J}{z-w} + \pa J\right)
\ee
\be
\overline J(z, \bar z) \overline J(w,\bar w) \sim \frac{1}{|z - w|^2}
\left(\frac{2 \overline J}{\bar z - \bar w}
+ \bar\pa \overline J\right)
\ee
\be
J(z, \bar z) \overline J(w,\bar w) \sim \frac{1}{|z - w|^2}
\left(\frac{S}{|z-w|^2} +
\frac{\pa S + J }{\bar z -\bar w}
+ \frac{\overline J}{z-w} + \pa \overline J \right).
\ee
\be
\overline J(z, \bar z) J(w,\bar w) \sim \frac{1}{|z - w|^2}
\left(\frac{S}{|z-w|^2} +
\frac{\bar\pa S + \overline J}{z - w}
+ \frac{J }{\bar z- \bar w} + \bar\pa J\right).
\ee

Turning now to the action of these operators on $\Phi$, we find
\be
S(z, \bar z) \Phi(w,\bar w) \sim  \frac{(h- \bar h)\Phi}{|z-w|^2}
\ee
\be
J(z, \bar z) \Phi(w,\bar w) \sim \frac{1}{|z - w|^2}
\left(\frac{ h \Phi}{z-w} + \pa \Phi\right)
\ee
\be
\overline J(z, \bar z) \Phi(w,\bar w) \sim \frac{1}{|z - w|^2}
\left(\frac{ \bar h \Phi}{\bar z -\bar w}
+ \bar\pa \Phi \right) .
\ee

\section{Implications for Celestial Holography}

The preceding analysis of diffeomorphism symmetry has an important implication
for celestial holography. In celestial holography the helicity of
a massless particle in four-dimensional spacetime is represented by the
two-dimensional conformal spin of the corresponding operator on the celestial
sphere. Thus, since the assumption of two-dimensional diffeomorphism symmetry
implies conservation of conformal spin, the holographic
implication is conservation of four-dimensional helicity in the interactions
of massless particles such
as gravitons and photons. This is definitely not what we want. Thus, we are
forced to conclude that the conformal symmetry of the holographic theory
on the celestial sphere must not be extended to diffeomorphism
symmetry.

To make perfectly clear what we are talking about, let us
consider a specific example -- the OPE of two positive-helicity graviton operators
derived in \cite{Pate:2019lpp}:
\be
G^+_{\D_1}(z_1, {\bar z}_1) G^+_{\D_2}(z_2, {\bar z}_2) \sim
E_+(\D_1,\D_2) \frac{{\bar z}_{12}}{z_{12}} G^+_{\D_1 + \D_2}(z_2, {\bar z}_2),
\ee
where $z_{12} = z_1 - z_2$ and
\be
E_+(\D_1,\D_2) = -\frac{\k}{2}B(\D_1 -1,\D_2-1).
\ee
The left-hand side of the OPE has conformal spin $s=2+2 =4$ and the right-hand side
has conformal spin
$s=2$. Thus, conformal spin is not conserved. As we have explained, this implies that
diffeomorphism symmetry is violated as well.

Celestial holography may require all
sorts of symmetries in addition to conformal symmetry, such as supertranslations and
a $w_{1+\infty}$ loop algebra \cite{Strominger:2021lvk},
but Diff($S^2$) is not one of them. There are
classical field theories that do conserve helicity, such as theories of
scalars only or Born--Infeld theory \cite{Cheung:2021zvb},
not coupled to gravity, but they are unlikely to be fully consistent as
quantum theories. Even if there were consistent helicity-conserving quantum theories, they
would not be expected to have a dual celestial realization with
infinite-dimensional conformal symmetry,
let alone diffeomorphism symmetry, since they do not contain gravity.

It may be interesting to make precise the implications of this conclusion
for the asymptotic metric analysis of GR in asymptotically flat
spacetime. Although diffeomorphism symmetry is not appropriate for celestial holography,
the description of the two-dimensional diffeomorphism
algebra presented here might be useful for other purposes.

\section*{Acknowledgments}

I am grateful to Clifford Cheung for reading a draft of this manuscript and
making helpful comments. This work was supported in part
by the Walter Burke Institute for Theoretical Physics at Caltech and by U.S. DOE
grant DE-SC0011632. It was performed in part at the Aspen Center for Physics,
which is supported by NSF grant PHY-1607611.

\end{document}